\documentclass[prd,preprint,superscriptaddress,showpacs,byrevtex]{revtex4}
\usepackage{bm}
\def\fsl#1{\setbox0=\hbox{$#1$}           
   \dimen0=\wd0                                 
   \setbox1=\hbox{/} \dimen1=\wd1               
   \ifdim\dimen0>\dimen1                        
      \rlap{\hbox to \dimen0{\hfil/\hfil}}      
      #1                                        
   \else                                        
      \rlap{\hbox to \dimen1{\hfil$#1$\hfil}}   
      /                                         
   \fi}                                         %
\newcommand{\be}{\begin{equation}}
\newcommand{\ee}{\end{equation}}
\newcommand{\bea}{\begin{eqnarray}}
\newcommand{\eea}{\end{eqnarray}}
\newcommand{\beq}{\begin{equation}}
\newcommand{\eeq}{\end{equation}}
\newcommand{\beqs}{\begin{eqnarray}}
\newcommand{\eeqs}{\end{eqnarray}}

\newcommand{\dslash}{D\hspace{-0.067in}\slash}

\begin{document}
\title{ Proof of Factorization of Fragmentation Function in Non-Equilibrium QCD }
\author{Gouranga C Nayak } \email{nayak@physics.arizona.edu}
\affiliation{ Department of Physics, University of Arizona, Tucson, AZ 85721, USA }
\begin{abstract}
In this paper we prove factorization of fragmentation function in non-equilibrium
QCD by using Schwinger-Keldysh closed-time path integral formalism. We use the
background field method of QCD in a pure gauge in path integral approach to prove factorization of
fragmentation function in non-equilibrium QCD. Our proof is valid in any arbitrary gauge fixing parameter
$\alpha$. This may be relevant to study hadron production from quark-gluon plasma at high energy heavy-ion
colliders at RHIC and LHC.
\end{abstract}
\maketitle
\pagestyle{plain}
\pagenumbering{arabic}
\section{Introduction}
RHIC and LHC heavy-ion colliders are the best facilities to study quark-gluon
plasma in the laboratory. Since two nuclei travel almost at speed of light,
the QCD matter formed at RHIC and LHC may be in non-equilibrium. In order
to make meaningful comparison of the theory with the experimental data on hadron
production, it may be necessary to study nonequilibrium-nonperturbative QCD at
RHIC and LHC. This, however, is a difficult problem.

Non-equilibrium quantum field theory can be studied by using Schwinger-Keldysh closed-time
path (CTP) formalism \cite{schw,keldysh}. However, implementing CTP in non-equilibrium at
RHIC and LHC is a very difficult problem, especially due to the presence of gluons in
non-equilibrium and hadronization etc. Recently, one-loop resummed gluon propagator in
non-equilibrium in covariant gauge is derived in \cite{greiner,cooper}.

High $p_T$ hadron production at high energy $e^+e^-$, $ep$ and $pp$ colliders is studied
by using Collins-Soper fragmentation function \cite{collins,sterman,george}.
For a high $p_T$ parton fragmenting to hadron, Collins-Soper derived an expression
for the fragmentation function based on field theory and factorization properties in QCD at
high energy. This fragmentation function is universal in the sense that, once
its value is determined from one experiment it explains the data at other experiments \cite{frag}.

Recently we have derived parton to hadron fragmentation function in non-equilibrium QCD
by using Schwinger-Keldysh closed-time path integral formalism \cite{nayak1}.
This can be relevant at RHIC and LHC heavy-ion colliders to study hadron production
from quark-gluon plasma. We have considered a high $p_T$ parton in medium at
initial time $\tau_0$ with arbitrary non-equilibrium (non-isotropic) distribution function
$f(\vec{p})$ fragmenting to hadron. The special case $f(\vec{p})=\frac{1}{e^{\frac{p_0}{T}}\pm 1}$
corresponds to the finite temperature QCD in equilibrium.

We have found the following definition of the parton to hadron fragmentation function
in non-equilibrium QCD by using closed-time path integral formalism.
For a quark ($q$) with arbitrary non-equilibrium distribution function $f_q(\vec{k})$ at
initial time, the quark to hadron fragmentation function is given by \cite{nayak1}
\bea
&& D_{H/q}(z,P_T)
= \frac{1}{2z[1+f_q({\vec k})]} \int dx^- \frac{d^{d-2}x_T}{(2\pi)^{d-1}}  e^{i{k}^+ x^- + i {P}_T \cdot x_T/z} \nonumber \\
 &&\frac{1}{2}{\rm tr_{Dirac}}\frac{1}{3}{\rm tr_{color}}[\gamma^+<in| \psi(x^-,x_T) \Phi^{-1}[x^-,x_T]a^\dagger_H(P^+,0_T)  a_H(P^+,0_T) \Phi[0]~{\bar \psi}(0)  |in>]
\label{qnf}
\eea
where $z$ (=$\frac{P^+}{k^+}$) is the longitudinal momentum fraction of the hadron with respect to the
parton and $P_T$ is the transverse momentum of the hadron.
In the above equations $|in>$ is the initial state of the non-equilibrium quark (gluon)
medium. The path ordered exponential
\bea
\Phi[x^\mu ]={\cal P}~ {\rm exp}[ig\int_{-\infty}^0 d\lambda~ n \cdot {\cal A}^a(x^\mu +n^\mu \lambda )~T^a]
\label{wilf}
\eea
is the Wilson line \cite{sterman,tucci}.

Eq. (\ref{qnf}) can be compared with the following definition of Collins-Soper
fragmentation function in vacuum \cite{collins}:
\bea
&& D_{H/q}(z,P_T)
= \frac{1}{2z} \int dx^- \frac{d^{d-2}x_T}{(2\pi)^{d-1}}  e^{i{k}^+ x^- + i {P}_T \cdot x_T/z} \nonumber \\
 &&\frac{1}{2}{\rm tr_{Dirac}}\frac{1}{3}{\rm tr_{color}}[\gamma^+<0| \psi(x^-,x_T) \Phi^{-1}[x^-,x_T]a^\dagger_H(P^+,0_T)  a_H(P^+,0_T) \Phi[0]{\bar \psi}(0)  |0>].
\label{qvf}
\eea
Eq. (\ref{qnf}) differs from eq. (\ref{qvf}) only by the presence of prefactor $\frac{1}{1+f_q({\vec k})}$ and
by the replacement of the vacuum $|0>$ by the state $|in>$.

Similar to the pp collisions, the $Q^2$ evolution of this non-equilibrium fragmentation function at RHIC and LHC
can be studied by using DGLAP evolution equation \cite{gl,ap,d} which we have recently extended to non-equilibrium
QCD \cite{nayakdglap}.

Another important physics analysis one needs to do at RHIC and LHC is to prove factorization of fragmentation function
in non-equilibrium QCD to study hadron production from quark-gluon plasma. Factorization refers to separation of short
distance from long distance effects in field theory.

In this paper we will prove factorization of fragmentation function in non-equilibrium QCD.
Factorization of fragmentation function in pp collisions is studied by Collins, Soper and Sterman by using
diagrammatic technique to all orders in perturbation theory. Recently, we have proved factorization of soft
and collinear divergences by using background field method of QCD \cite{nayak2} in a pure gauge in path
integral approach. We have used the gauge fixing identity
\cite{nayakgf} which relates the generating functional $Z[A,J,\eta,{\bar \eta}]$
in the background field method of QCD in pure gauge with $Z[J,\eta,{\bar \eta}]$ in QCD (without the background field).

We find it convenient to use background field method in a pure gauge to prove factorization of soft and collinear divergences
in non-equilibrium quantum field theory \cite{nayakqed}.
In this paper we will extend this to non-equilibrium QCD. Unlike QED \cite{tucci}, the gauge fixing term and
ghost term in the background field method of QCD \cite{thooft,abbott} depend on the background field $A_\mu^a(x)$.
Hence the extension of proof of factorization from non-equilibrium QED to non-equilibrium QCD is not straightforward.
For this reason we will use the gauge fixing identity \cite{nayakgf} to prove factorization of fragmentation function
in non-equilibrium QCD by using  Schwinger-Keldysh closed-time path
integral formalism in background field method of QCD in a pure gauge.
Our proof is valid in any arbitrary gauge fixing parameter $\alpha$. This can be relevant
to study hadron production from quark-gluon plasma at RHIC and LHC.

The paper is organized as follows. In section II we briefly review the factorization of soft and
collinear divergences in QCD in pp collisions by using background field method of QCD. In section III we
describe Schwinger-Keldysh closed-time path integral formalism in non-equilibrium QCD relevant for
our purpose. In section IV we prove factorization of fragmentation function in non-equilibrium QCD
by using background field method of QCD and closed-time path integral formalism. Section V contains
conclusions.

\section{ Background Field Method of QCD and Factorization of Soft and Collinear Divergences }

Background field method of QCD was originally formulated by 't Hooft \cite{thooft} and later
extended by Abbott \cite{abbott}. This is an elegant formalism which can be useful to construct gauge invariant
(off-shell) green's functions in QCD. This is because, unlike QCD, the lagrangian density in the generating
functional in the background field method of QCD is background gauge invariant (even after quantizing the theory).
In the presence of external sources a relation between the generating functional $Z[J,\eta, {\bar \eta}]$ in QCD and
the generating functional $Z[A,J,\eta, {\bar \eta}]$ in the background field method of QCD in pure gauge
is recently obtained \cite{nayakgf}. We will use this relation to prove factorization of fragmentation function
in non-equilibrium QCD.

In QCD, the generating functional is given by
\bea
Z_{\rm QCD}[J,\eta,{\bar \eta}]=\int [dQ] [d{\bar \psi}] [d \psi ] ~{\rm det}(\frac{\delta \partial_\mu Q^{\mu a}}{\delta \omega^b})
~e^{i\int d^4x [-\frac{1}{4}{F^a}_{\mu \nu}^2[Q] -\frac{1}{2 \alpha} (\partial_\mu Q^{\mu a})^2+{\bar \psi} \dslash [Q] \psi + J \cdot Q +{\bar \eta} \psi + \eta  {\bar \psi} ]}. \nonumber \\
\label{zfq}
\eea
Under the infinitesimal gauge transformation the quantum gluon field transforms as
\bea
\delta Q_\mu^a = -gf^{abc}\omega^b Q_\mu^c + \partial_\mu \omega^a
\label{lambda}
\eea
where $\omega^a$ is the gauge transformation parameter. In the background field method of QCD the
generating functional is given by \cite{thooft,abbott}
\bea
&& Z_{\rm background ~QCD}[A,J,\eta,{\bar \eta}]=\int [dQ] [d{\bar \psi}] [d \psi ] ~{\rm det}(\frac{\delta G^a(Q)}{\delta \omega^b}) \nonumber \\
&& e^{i\int d^4x [-\frac{1}{4}{F^a}_{\mu \nu}^2[A+Q] -\frac{1}{2 \alpha}
(G^a(Q))^2+{\bar \psi} \dslash [A+Q] \psi + J \cdot Q +{\bar \eta} \psi + \eta {\bar \psi} ]}
\label{zaqcd}
\eea
where $A_\mu^a$ is the background field. The gauge fixing term is given by
\bea
G^a(Q) =\partial_\mu Q^{\mu a} + gf^{abc} A_\mu^b Q^{\mu c}=D_\mu[A]Q^{\mu a}
\label{ga}
\eea
which depends on the background field $A_\mu^a$. The infinitesimal gauge transformation is given by \cite{abbott}
\bea
&& \delta Q_\mu^a = -gf^{abc}\omega^b (A_\mu^c +Q_\mu^c) + \partial_\mu \omega^a \nonumber \\
&& \delta A_\mu^a =0.
\label{omega1}
\eea
Changing the variable $Q \rightarrow Q-A$ in eq. (\ref{zaqcd}) we find
\bea
&& Z_{\rm background ~QCD}[A,J,\eta,{\bar \eta}]= e^{-i\int d^4x J \cdot A}~ \int [dQ] [d{\bar \psi}] [d \psi ] ~{\rm det}(\frac{\delta G_f^a(Q)}{\delta \omega^b}) \nonumber \\
&& e^{i\int d^4x [-\frac{1}{4}{F^a}_{\mu \nu}^2[Q] -\frac{1}{2 \alpha} (G_f^a(Q))^2+{\bar \psi} \dslash [Q] \psi + J \cdot Q + \eta {\bar \psi}
+{\bar \eta} \psi ]}
\label{zaqcd1}
\eea
where
\bea
G_f^a(Q) =\partial_\mu Q^{\mu a} + gf^{abc} A_\mu^b Q^{\mu c} - \partial_\mu A^{\mu a}=D_\mu[A] Q^{\mu a} - \partial_\mu A^{\mu a}.
\label{gfa}
\eea

The most crucial statement of factorization theorem of Collins, Soper and Sterman is the appearance of the
Wilson line eq. (\ref{wilf}) in the definition of the parton distribution function \cite{collins}
\bea
&& f_{q/P}(x,k_T)
= \frac{1}{2}~\int dy^- \frac{d^{d-2}y_T}{(2\pi)^{d-1}}  e^{ix{P}^+ y^- + i {k}_T \cdot y_T} \nonumber \\
&&~\frac{1}{2}{\rm tr_{Dirac}}~\frac{1}{3}{\rm tr_{color}}[\gamma^+<P| {\bar \psi}(y^-,y_T) ~\Phi[y^-,y_T] ~\Phi^{-1}[0]~\psi(0)  |P>].
\eea
and in fragmentation function in eq. (\ref{qvf}) which makes them gauge invariant. This Wilson line is responsible for
cancelation of soft and collinear divergences which arise due to presence of loops and/or higher order Feynman
diagrams. This Wilson line can be thought of as a quark or gluon jet propagating in a soft and/or collinear gluon
cloud. The Wilson line eq. (\ref{wilf}) can be written as
\bea
&&~\Phi[x^\mu ]={\cal P}~ {\rm exp}[ig\int_{-\infty}^0 d\lambda~ h \cdot {\cal A}^a(x^\mu +h^\mu \lambda )~T^a] \nonumber \\
&& ={\cal P}~ {\rm exp}[ig\int_{-\infty}^0 d\lambda~ h \cdot e^{\lambda h \cdot \partial} {\cal A}^a(x^\mu )~T^a]
={\cal P}~ {\rm exp}[ig \frac{1}{ h \cdot \partial} h \cdot {\cal A}^a(x^\mu )~T^a].
\label{wilf1}
\eea
Since the vector $h^\mu$ is free we can choose it to correspond to various physical situations.
For example if we choose $h^\mu=n^\mu$, where $n^\mu$ is a fixed lightlike vector, we find from eq. (\ref{wilf1}) the phase factor
\bea
\omega^a(x)= \frac{1}{ n \cdot \partial} n \cdot  {\cal A}^a(x ).
\label{eik}
\eea
Using the Fourier transformation
\bea
{\cal A}^a_\mu(x) =\int \frac{d^4k}{(2\pi)^4} {\cal A}^a_\mu(k) e^{ik \cdot x}
\label{ft}
\eea
we find from eq. (\ref{eik})
\bea
V=g~\omega(k)=ig~\frac{n^\mu}{n \cdot k}.
\label{eikonal}
\eea
Note that eq. (\ref{eikonal}) is precisely the Eikonal vertex for a soft gluon with momentum $k$ interacting with a
high energy quark (or gluon) jet moving along the direction $n^\mu$ \cite{collins,sterman}. In the soft gluon
approximation in \cite{sterman} the fixed lightlike vector is taken to be having only "+" or "-" component:
\bea
n^\mu=(n^+,n^-,n_T)=(1,0,0)~~~~~~~~~~~{\rm or}~~~~~~~~~n^\mu=(n^+,n^-,n_T)=(0,1,0).
\label{n}
\eea

Now we will show that when the classical background field is an abelian-like pure gauge
\bea
{\cal A}_\mu^a (x) =\partial_\mu \omega^a(x)
\label{pure}
\eea
we can reproduce eqs. (\ref{eik}) and (\ref{eikonal}) which appear in the Wilson line eq. (\ref{wilf1}).
Multiplying $x^\mu$ independent four vector $h^\mu$ from left in eq. (\ref{pure}) we find
\bea
h \cdot {\cal A}^a(x) = h \cdot \partial \omega^a(x).
\label{om2}
\eea
Dividing $h \cdot \partial $ from left in the above equation we find
\bea
\omega^a(x) = \frac{1}{h \cdot \partial } h \cdot {\cal A}^a(x).
\label{om3}
\eea
Since the four vector $h^\mu$ is free we can choose it such way that ${\cal A}_\mu^a$ can represent Feynman rules
involving soft gluons or collinear gluons. For example, when $h^\mu =n^\mu$ where $n^\mu$ is a fixed light like vector (see
eq. (\ref{n})), we find from the above equation
\bea
\omega^a(x) = \frac{1}{n \cdot \partial } n \cdot {\cal A}^a(x)
\label{om4}
\eea
which reproduces eq. (\ref{eik}) which appears inside the Wilson line in eq. (\ref{wilf1}).

Similarly, if we choose $h^\mu=n^\mu_B$, where $n^\mu_B$ is a non-light like vector
\bea
n^\mu_B=(n^+_B,n^-_B,0),
\label{nb}
\eea
we reproduce the Feynman rules for the collinear divergences \cite{sterman}.
This establishes the correspondence between the Wilson line and the classical background field ${\cal A}_\mu^a(x)$
as given by eq. (\ref{pure}) in the context of soft and/or collinear divergences.

The non-perturbative correlation function in the presence of the background field $A_\mu^a(x)$ is given by
\bea
\frac{\delta}{\delta {\bar \eta}_i(x_2) }~\frac{\delta}{\delta { \eta}_j(x_1) }
~Z_{\rm background ~QCD}[A,J,\eta,{\bar \eta}]|_{J=\eta={\bar \eta}=0}~~=~~<\psi_i(x_2) {\bar \psi}_j(x_1)>^A_{\rm background~QCD}.
\label{g1q}
\eea
Similarly the non-perturbative correlation function in QCD (without the background field) is given by
\bea
\frac{\delta}{\delta {\bar \eta}_i(x_2) }~\frac{\delta}{\delta { \eta}_j(x_1) }
~Z_{\rm QCD}[J,\eta,{\bar \eta}]|_{J=\eta={\bar \eta}=0}~~=~~<\psi_i(x_2) {\bar \psi}_j(x_1)>^{A=0}_{\rm QCD}.
\label{g2q}
\eea

When the background field $A_\mu^a(x)$ is SU(3) pure gauge in QCD given by
\bea
T^aA_\mu^a=\frac{1}{ig} (\partial_\mu U)U^{-1},~~~~~~~~~~~~~~U=e^{igT^a\omega^a(x)}
\label{su3pure}
\eea
we find from \cite{nayak2}
\bea
&& <\psi_i(x_2) {\bar \psi}_j(x_1)>^A_{\rm background~QCD}~=
 ~[{\cal P}~ {\rm exp}[ig\int_{-\infty}^{0} d\lambda~ h \cdot {\cal A}^a(x^\mu_2 +h^\mu \lambda )T^a]] \nonumber \\
&&~\times ~[<\psi_i(x_2) {\bar \psi}_j(x_1)>^{A=0}_{\rm QCD}]~\times ~[{\bar {\cal P}}~ {\rm exp}[ig\int_0^{-\infty} d\lambda~ h \cdot {\cal A}^b(x^\nu_1 +h^\nu \lambda )T^b]].
\label{fact1f}
\eea
All the ${\cal A}_\mu^a(x)$ dependences (responsible for soft and collinear divergences)
have been factored into the path ordered exponentials or Wilson lines. This proves factorization
of soft and collinear divergences by using background field method of QCD.

Note that the structure function and fragmentation function
are proportional to the non-perturbative correlation function in eq. (\ref{fact1f}) with the following identification.
In the generating functional $Z[A,J,\eta, {\bar \eta}]$ in the background field method of QCD
the integration is only over those functions of quantum gluon field $Q_\mu^a(x)$, quark field $\psi_i(x)$ and
antiquark field ${\bar \psi}_i(x)$ such that their Fourier transform $Q_\mu^a(k)$, $\psi_i(k)$ and ${\bar \psi}_i(k)$
vanish in the soft region \cite{tucci}.
\section{ Closed-Time Path Integral Formalism in Non-Equilibrium QCD }
Unlike $pp$ collisions, the ground state at RHIC and LHC heavy-ion collisions
(due to the presence of a QCD medium at initial time $t=t_{in}$ (say $t_{in}$=0)
is not a vacuum state $|0>$ any more. We denote $|in>$ as the initial state of the non-equilibrium QCD
medium at $t_{in}$.

Consider the time evolution of the density matrix
\bea
i\frac{\partial \rho}{\partial t} = [H_I, \rho],~~~~~~~~~~~~~~~~\rho(-\infty) = \rho_0
\label{rho1}
\eea
where $H_I$ is the interaction hamiltonian. The formal solution is
\bea
\rho(t)=S(t,-\infty)\rho_0 S(-\infty,t),~~~~~~~~~~~~~~~~~~~S(t,-\infty)=T~exp[-i\int_{-\infty}^t dt'~H_I(t')].
\label{rho2}
\eea
The density matrix $\rho$ is in interaction picture. The average value of an operator $L$ in the interaction
picture is given by
\bea
<L(t)>=Tr[\rho(t)L(t)],~~~~~~~~~~~~~~~~~~~~~~~~~~i\frac{\partial L(t)}{\partial t} = [H_0, L(t)]
\label{rho3}
\eea
where $H_0$ is the free hamiltonian. Since in many situations we deal with correlation function of several fields
at different times, it is useful to transfer all the time dependence to field operators and consider the density
operator as independent of time, {\it i.e.} to go to Heisenberg representation. For the time independence of the density
matrix we can take the value of the matrix determined by expression eq. (\ref{rho2}) at a certain fixed instant
of time, for example, $t=t_{in}=0$, having thus included in it all the changes which the distribution $\rho_0$ had
undergone when the external field and the interaction in the system were switched on.

Consider the medium average of $T-$products of several operators. Using the Heisenberg density matrix one finds
\bea
&& <T[L(t)M(t')...]>=Tr[\rho T[L(t)M(t')....]]=Tr[ S(0,-\infty)\rho_0 S(-\infty,0)T[L(t)M(t')....]] \nonumber \\
&& =Tr[ \rho_0 S(-\infty,0)T[L(t)M(t')....]S(0,-\infty)].
\label{rho4}
\eea
Going over to the operators in the interaction picture
\bea
&&<T[L(t)M(t')...]>=Tr[ \rho_0 S(-\infty,0)T[S(0,t)L_I(t)S(t,t')M_I(t')....]S(0,-\infty)] \nonumber \\
&&=Tr[ \rho_0 T_c[S_cL_I(t)M_I(t')....]]
\label{rho5}
\eea
where $T_c$ is the complete contour from
\bea
-\infty \rightarrow t \rightarrow t' ....\rightarrow -\infty
\label{rho6}
\eea
and $S_c$ is the complete $S-$matrix defined along $T_c$.

To deal with Feynman diagram and Wick theorem it is useful to split the time interval to "+" and "-" contour;
where "+" time branch is from $-\infty$ to $+\infty$ where (time) $T-$order product apply and "-" time branch
is from $+\infty$ to $-\infty$ where (anti-time)${\bar T}-$order product apply.

Consider scalar field theory first. Since there are two time
branches there are two fields and two sources and hence four Green's functions.
Let us denote the field $\phi^+(x)$ and the source $J^+(x)$ in the "+" time branch
and $\phi^-(x)$ and $J^-(x)$ in the "-" time branch. The generating functional is given by
\bea
Z[\rho,J_+,J_-]=\int [d\phi_+][d\phi_-] ~{\rm exp}[i[S[\phi_+]-S[\phi_-]+\int d^4x J_+\phi_+-\int d^4x J_-\phi_-]]~<\phi_+,0|\rho|0,\phi_-> \nonumber \\
\label{rho7}
\eea
where $S[\phi]$ is the full action in scaler field theory and $|\phi_{\pm},0>$ is the quantum state corresponding to the field
configuration $\phi_{\pm}(\vec{x},t=0)$.

In the CTP formalism in non-equilibrium there are four Green's functions
\bea
&& G_{++}(x,x') = \frac{\delta Z[J_+,J_-,\rho]}{i^2 \delta J_+(x) J_+(x')}=<in|T\phi (x) \phi (x')|in> = <T\phi (x) \phi (x')>\nonumber \\
&& G_{--}(x,x') = \frac{\delta Z[J_+,J_-,\rho]}{(-i)^2 \delta J_-(x) J_-(x')}= <in|{\bar T} \phi (x) \phi (x')|in> = <{\bar T} \phi (x) \phi (x')>\nonumber \\
&& G_{+-}(x,x') = \frac{\delta Z[J_+,J_-,\rho]}{-i^2 \delta J_+(x) J_-(x')}= <in|\phi (x') \phi (x)|in> = <\phi (x') \phi (x) >\nonumber \\
&& G_{-+}(x,x') = \frac{\delta Z[J_+,J_-,\rho]}{-i^2 \delta J_-(x) J_+(x')}= <in|\phi (x) \phi (x')|in>= <\phi (x) \phi (x') >
\label{green}
\eea
where $T$ is the time order product and ${\bar T}$ is the anti-time order product given by
\bea
&& T\phi (x) \phi (x') = \theta(t - t') \phi (x) \phi (x') + \theta(t'-t) \phi (x') \phi (x) \nonumber \\
&& {\bar T} \phi (x) \phi (x') = \theta(t' - t) \phi (x) \phi (x') + \theta(t-t') \phi (x') \phi (x).
\label{ttbar}
\eea

\subsection{Generating Functional in Non-Equilibrium QCD}

Additional ghost fields are present in the gauge theory. However, we will directly work with the determinant of the
gauge fixing terms. The generating functional in non-equilibrium QCD is given by
\bea
&& Z[\rho,J_+,J_-,\eta_+,\eta_-,{\bar \eta}_+,{\bar \eta}_-]=\int [dQ_+] [dQ_-][d{\bar \psi}_+] [d{\bar \psi}_-] [d \psi_+ ] [d\psi_-]~
\times \nonumber \\
&& \times ~{\rm det}(\frac{\delta \partial_\mu Q_+^{\mu a}}{\delta \omega_+^b})~\times ~{\rm det}(\frac{\delta \partial_\mu Q_-^{\mu a}}{\delta \omega_-^b}) ~\times \nonumber \\
&& {\rm exp}[i\int d^4x [-\frac{1}{4}({F^a}_{\mu \nu}^2[Q_+]-{F^a}_{\mu \nu}^2[Q_-])-\frac{1}{2 \alpha} (
(\partial_\mu Q_+^{\mu a })^2-(\partial_\mu Q_-^{\mu a })^2) +{\bar \psi}_+ \dslash [Q_+] \psi_+ \nonumber \\
&& -{\bar \psi}_- \dslash [Q_-] \psi_-
+ J_+ \cdot Q_+ -J_- \cdot Q_-+{\bar \eta}_+ \cdot \psi_+ - {\bar \eta}_- \cdot \psi_- + \eta_+ \cdot {\bar \psi}_+
- \eta_- \cdot {\bar \psi}_-]]
\nonumber \\
&& \times ~<Q_+,\psi_+,{\bar \psi}_+,0|~\rho~|0,{\bar \psi}_-,\psi_-,Q_->
\label{zfqinon}
\eea
where $\rho$ is the initial density of state. The state $|Q^\pm,\psi^\pm,{\bar \psi}^\pm,0>$ corresponds to the field
configurations $Q_\mu^a({\vec x},t=t_{in}=0)$, $\psi({\vec x},t=t_{in}=0)$ and ${\bar \psi}({\vec x},t=t_{in}=0)$ respectively.
Note that we work in the frozen ghost formalism \cite{greiner,cooper} for the medium part at the initial time $t=t_{in}=0$.
\section{ Proof of Factorization in Non-Equilibrium QCD }
The generating functional in the background field method of QCD is given by eq. (\ref{zaqcd1}) where the gauge fixing term $G_f(Q)$ is
given by eq. (\ref{gfa}). Now combining the background field method of QCD with the closed-time path integral formalism we find
the generating functional in non-equilibrium QCD in the presence of background field $A_\mu^a$
\bea
&& Z_{\rm background ~QCD}[\rho,A,J_+,J_-,\eta_+,\eta_-,{\bar \eta}_+,{\bar \eta}_-]= e^{-i\int d^4x (J_+ \cdot A_+-J_- \cdot A_-)} \nonumber \\
&&~\int [dQ_+] [dQ_-][d{\bar \psi}_+] [d{\bar \psi}_-] [d \psi_+ ] [d\psi_-]~\times ~{\rm det}(\frac{\delta G_f^a(Q_+)}{\delta \omega_+^b})~\times ~{\rm det}(\frac{\delta G_f^a(Q_-)}{\delta \omega_-^b}) \times \nonumber \\
&& {\rm exp}[i\int d^4x [-\frac{1}{4}({F^a}_{\mu \nu}^2[Q_+]-{F^a}_{\mu \nu}^2[Q_-])-\frac{1}{2 \alpha} (
((G_f^a(Q_+))^2-(G_f^a(Q_-))^2) +{\bar \psi}_+ \dslash [Q_+] \psi_+ \nonumber \\
&& -{\bar \psi}_- \dslash [Q_-] \psi_-
+ J_+ \cdot Q_+ -J_- \cdot Q_-+{\bar \eta}_+ \cdot \psi_+ - {\bar \eta}_- \cdot \psi_- + \eta_+ \cdot {\bar \psi}_+
- \eta_- \cdot {\bar \psi}_-]]
\nonumber \\
&& \times ~<Q_+,\psi_+,{\bar \psi}_+,0|~\rho~|0,{\bar \psi}_-,\psi_-,Q_->.
\label{backnon1}
\eea
The fermion fields and corresponding sources transform as follows
\bea
&& \psi'_+ = U_+~ \psi_+,~~~~~~~~~{\bar \psi}'_+ = {\bar \psi}_+ ~U_+^{-1},~~~~~~~~~~~~~~\eta'_+ = U_+~ \eta_+,~~~~~~~~~{\bar \eta}'_+ =
{\bar \eta}_+ U^{-1}_+ \nonumber \\
&& \psi'_- = U_-~ \psi_-,~~~~~~~~~\bar \psi'_- = {\bar \psi}_- ~U_-^{-1},~~~~~~~~~~~~~~\eta'_- = U_-~ \eta_-,~~~~~~~~~{\bar \eta}'_- =
{\bar \eta}_-~ U^{-1}_-
\label{ftnn}
\eea
where
\bea
U_+=e^{igT^a\omega^a_+(x)},~~~~~~~~~~~~~~~U_-=e^{ig T^a \omega^a_-(x)}.
\label{purepm}
\eea
When the background field $A_\mu^a(x)$ is pure gauge in QCD given by eq. (\ref{su3pure}) we find from \cite{nayak2}
\bea
&& Z_{\rm QCD}[J,\eta,{\bar \eta}] ~=e^{i\int d^4x J \cdot A}~\times~Z_{\rm background ~QCD}[A,J,\eta',{\bar \eta}']
-~\int [dQ] [d{\bar \psi}] [d \psi ]~{\rm det}(\frac{\delta G_f^a(Q)}{\delta \omega^b}) \nonumber \\
&&~{\rm exp}[i\int d^4x [-\frac{1}{4}{F^a}_{\mu \nu}^2[Q] -\frac{1}{2 \alpha} (G_f^a(Q))^2+{\bar \psi} \dslash [Q] \psi + J \cdot Q +{\bar \eta} \psi
+  \eta {\bar \psi} ]] ~\times~[i \int d^4x  J \cdot \delta Q+...] \nonumber \\
\label{finalq}
\eea
where we have used the gauge fixing identity \cite{nayakgf}. Since we work in the frozen ghost formalism \cite{greiner,cooper}
for the medium part at the initial time $t=t_{in}=0$, the initial state $|{\bar \psi}_-,\psi_-,Q_-,0>$
associated with the medium part at initial time is gauge invariant
by construction. Hence we find by extending eq. (\ref{finalq}) to non-equilibrium QCD
\bea
&& Z_{\rm QCD}[\rho,J_+,J_-,\eta_+,\eta_-,{\bar \eta}_+,{\bar \eta}_-] ~=e^{i\int d^4x [J_+ \cdot A_+ - J_- \cdot A_- ]}~\times~Z_{\rm background ~QCD}[\rho,A,J_+,J_-,\eta'_+,\eta'_-,{\bar \eta}'_+,{\bar \eta}'_-] \nonumber \\
&& -\int [dQ_+][dQ_-] [d{\bar \psi}_+] [d{\bar \psi}_-] [d \psi_+ ][d \psi_- ]~\times ~{\rm det}(\frac{\delta G_f^a(Q_+)}{\delta \omega_+^b}) ~\times ~{\rm det}(\frac{\delta G_f^a(Q_-)}{\delta \omega_-^b}) ~\times \nonumber \\
&& \times ~[i \int d^4x ~ [J_+ \cdot \delta Q_+-J_- \cdot \delta Q_-]+...] ~\times \nonumber \\
&&~{\rm exp}[i\int d^4x [-\frac{1}{4}({F^a}_{\mu \nu}^2[Q_+]-{F^a}_{\mu \nu}^2[Q_-])
-\frac{1}{2 \alpha} ((G_f^a(Q_+))^2-(G_f^a(Q_-))^2)+{\bar \psi}_+ \dslash [Q_+] \psi_+ \nonumber \\
&& -{\bar \psi}_- \dslash [Q_-] \psi_-  + J_+ \cdot Q_+-J_- \cdot Q_- +{\bar \eta}_+ \psi_+ -{\bar \eta}_- \psi_-+  \eta_+ {\bar \psi}_+
- \eta_- {\bar \psi}_-]] \nonumber \\
&& ~\times ~<Q_+,\psi_+,{\bar \psi}_+,0|~\rho~|0,{\bar \psi}_-,\psi_-,Q_->.
\label{finalqn}
\eea

The non-perturbative correlation function in the presence of the background field $A_\mu^a(x)$ is given by
\bea
&&~\frac{\delta}{\delta {\bar \eta}^i_r(x_2) }~\frac{\delta}{\delta { \eta}^j_s(x_1) }
~Z_{\rm background ~QCD}[A,J_+,J_-,\eta_+,\eta_-,{\bar \eta}_+,{\bar \eta}_-]|_{J_+=J_-=\eta_+={\bar \eta}_+=\eta_-={\bar \eta}_-=0}\nonumber \\
&& =<\psi^i_r(x_2) {\bar \psi}^j_s(x_1)>^A_{\rm background~QCD}=<in|\psi^i_r(x_2) {\bar \psi}^j_s(x_1)|in>^A_{\rm background~QCD}
\label{g1qn}
\eea
where $r,s=+,-$ are the closed-time indices. The non-perturbative correlation function in QCD (without the background field) is given by
\bea
&&~\frac{\delta}{\delta {\bar \eta}^i_r(x_2) }~\frac{\delta}{\delta { \eta}^j_s(x_1) }
~Z_{\rm QCD}[J_+,J_-,\eta_+,\eta_-,{\bar \eta}_+,{\bar \eta}_-]|_{J_+=J_-=\eta_+={\bar \eta}_+=\eta_-={\bar \eta}_-=0} \nonumber \\
&&=<\psi^i_r(x_2) {\bar \psi}^j_s(x_1)>^{A=0}_{\rm QCD}=<in|\psi^i_r(x_2) {\bar \psi}^j_s(x_1)|in>^{A=0}_{\rm QCD}.
\label{g2qn}
\eea
Using eq. (\ref{finalqn}) we find from eqs. (\ref{g1qn}) and (\ref{g2qn})
\bea
&& <in|\psi^i_r(x_2) {\bar \psi}^j_s(x_1)|in>^A_{\rm background~QCD} ~= \nonumber \\
&& ~[{\cal P}~e^{ig T^a\omega^a_r(x_2)}] ~\times ~[<in|\psi^i_r(x_2) {\bar \psi}^j_s(x_1)|in>^{A=0}_{\rm QCD}]~\times ~[{\bar {\cal P}}~ e^{-ig T^b \omega^b_s(x_1)}]
\label{fact1fnpp}
\eea
which gives by using eqs. (\ref{om3}) and (\ref{wilf1})
\bea
&& <in|\psi^i_r(x_2) {\bar \psi}^j_s(x_1)|in>^A_{\rm background~QCD} ~=~
 ~[{\cal P}~ {\rm exp}[ig\int_{-\infty}^{0} d\lambda~ h \cdot {\cal A}_r^a(x^\mu_2 +h^\mu \lambda )T^a]] \nonumber \\
&&~\times ~[<in|\psi^i_r(x_2) {\bar \psi}^j_s(x_1)|in>^{A=0}_{\rm QCD}]~\times ~[{\bar {\cal P}}~ {\rm exp}[ig\int_0^{-\infty} d\lambda~ h \cdot {\cal A}_s^b(x^\nu_1 +h^\nu \lambda )T^b]]
\label{fact1fn}
\eea
where the closed-time path indices $r,s$ are not summed.
All the ${\cal A}_\mu^a(x)$ dependences (responsible for soft and collinear divergences) have been factored
into the path ordered exponentials or Wilson lines. This proves factorization of soft and collinear
divergences in non-equilibrium QCD.
\section{ Conclusions }
In order to study hadron production from quark-gluon plasma using the non-equilibrium fragmentation
function one needs to prove factorization of fragmentation function in non-equilibrium QCD. Factorization refers to separation
of short distance from long distance effects in field theory. The factorization of soft and collinear divergences
in non-equilibrium QED is proved in \cite{nayakqed}. In this paper we have proved factorization of
fragmentation function in non-equilibrium QCD by using Schwinger-Keldysh closed-time
path integral formalism and background field method of QCD in a pure gauge. Our proof is valid in any arbitrary gauge
fixing parameter $\alpha$. This may be relevant to study hadron production \cite{nayak1} from quark-gluon plasma
\cite{qgp} at RHIC and LHC.

\acknowledgements

This work was supported in part by Department of Energy under contracts DE-FG02-91ER40664,
DE-FG02-04ER41319 and DE-FG02-04ER41298.

\end{document}